\documentclass[preprint,aps,prd,nofootinbib,tightenlines]{revtex4}
\usepackage{epsfig,graphicx,amsmath,amssymb,mathrsfs,float,color,xspace,helvet}
\definecolor{dblue}{rgb}{0.00,0.00,0.75}
\usepackage[colorlinks,urlcolor=dblue,linkcolor=dblue,citecolor=dblue]{hyperref}
\begin{document}
\title{Resonances $\rho(1450)^+$ and $\rho(1700)^+$ in $B \to D KK$ decays}

\author{Ai-Jun Ma$^1$}        \email{theoma@163.com}
\author{Wen-Fei Wang$^2$} \email{wfwang@sxu.edu.cn}

\affiliation{$^1$Department of Mathematics and Physics,  Nanjing Institute of Technology, Nanjing, Jiangsu 211167, P.R. China \\
                 $^2$Institute of Theoretical Physics, Shanxi University, Taiyuan, Shanxi 030006, P.R. China  }
\date{\today}
\begin{abstract}
The contributions for the kaon pair from the intermediate states $\rho(1450)^+$ and $\rho(1700)^+$ in the decays
$B^+ \to \bar{D}^0 K^+ \bar{K}^0$, $B^0  \to D^- K^+ \bar{K}^0$, and $B_s^0 \to  D_s^-K^+ \bar{K}^0$ are analyzed
within the perturbative QCD factorization approach. The decay amplitudes for all concerned decays in this work are
dominated by the factorizable Feynman diagrams with the emission of the kaon pair, and the charged $\rho$ mesons
should be of great importance in the $KK$ channel of the related three-body $B$ decays. Moreover, these quasi-two-body
decays are CKM-favored, and the relevant branching ratios are predicted to be in the order of $10^{-5}$, which have the potential to be
measured by experiments. It is also shown that the contributions of the subprocesses $\rho(1450, 1700)^+ \to KK$ for the
three-body $B$ meson decays are considerable according to the total three-body branching fractions presented by Belle.
Therefore, the decays $B^+ \to \bar{D}^0 K^+ \bar{K}^0$, $B^0  \to D^- K^+ \bar{K}^0$, and $B_s^0 \to  D_s^-K^+ \bar{K}^0$
can be employed to study the  properties of $\rho(1450)$ and $\rho(1700)$ in the LHCb and Belle-II experiments.
\end{abstract}


\maketitle

\section{INTRODUCTION}\label{sec-int}

Three-body hadronic $B$ meson decays provide us rich opportunities to explore \emph{CP} violations in the relevant
decay channels and to study intermediate resonance structures for the final states. Resonance contributions for the kaon
pair in three-body $B$ decays have been extensively studied experimentally~\cite{plb542-171,
prd71-092003,prd74-032003,prl100-171803,prd85-112010,
prd85-054023,prd87-072004,prd88-072005,jhep08-037,prd98-071103,jhep01-131} and theoretically~\cite{prd67-034012,
prd72-094003,plb622-207,prd76-094006,plb699-102,plb737-201,prd89-094007,epjc75-609,prd94-094015,prd95-036013,
prd96-113003,prd99-093007,prd102-056017,epjc80-394,epjc80-517,epjc80-815,ijmpa35-2050164} in recent years.
Nevertheless, the contributions from $\rho(770)$ and its excited states $\rho(1450)$ and $\rho(1700)$ for the kaon pair in the final
states of these decays were considered to be small and have been ignored in most cases.

Recently, an unexpected fit fraction of approximately $30\%$ for the $K^+K^-$ pair from resonance $\rho(1450)$ in the decays
$B^\pm \to \pi^\pm K^+K^-$was reported by the LHCb Collaboration in Ref.~\cite{prl123-231802}. This fit fraction is unreliable when
it is compared with the proportion contributed by the same state $\rho(1450)$ for the pion pair in $B^\pm \to \pi^\pm \pi^+\pi^-$
decays~\cite{prd101-012006,prl124-031801}. In~\cite{prd101-111901}, the perturbative QCD
(PQCD)~\cite{plb504-6,prd63-054008,prd63-074009,ppnp51-85} prediction for the branching fraction of
$B^\pm \to \pi^\pm \rho(1450)^0 \to \pi^\pm K^+K^-$ is much smaller than  LHCb's result in~\cite{prl123-231802},
even taking the contribution from the Breit-Wigner (BW) tail of $\rho(770)$ into account. The calculation within the light-cone sum rule
approach~\cite{2111.05647} does not support the result for $\rho(1450)^0\to K^+K^-$ from LHCb~\cite{prl123-231802} either.
In Ref.~\cite{prd103-013005}, the authors claimed the unsettled state $f_X(1500)$ is the vector $\rho(1450)$ reported by LHCb.
A broad resonance $\rho_X(1500)$ has also been promoted to describe this broad resonance  around the mass region of
$1.5~{\rm GeV}$~\cite{prd102-053006}.

The vector resonance $\rho(770)$ has been well established. However the properties of its excited states $\rho(1450)$ and $\rho(1700)$
are not yet entirely clear. It has been shown that the neutral intermediate resonances $\rho(1450)^0$ and $\rho(1700)^0$ contribute
the $K\bar{K}$ pair in the processes $\bar{p} p \to K^+K^-\pi^0$~\cite{plb468-178,plb639-165,
epjc80-453}, $e^+e^- \to K^+K^-$~\cite{pl99b-257,pl107b-297,prd76-072012,prd88-032013,prd94-112006,
plb779-64,zpc39-13}, $e^+ e^- \to K^0_{S}K^0_{L}$~\cite{pl99b-261,prd63-072002,plb551-27,plb760-314,
prd89-092002}, and $J/\psi \to K^+K^-\pi^0$~\cite{prl97-142002,prd75-074017,prd76-094016,prd95-072007,prd100-032004}.
To capture the information on $\rho(1450)$ and $\rho(1700)$ clearly, the investigation on the charged excited rho mesons in
$K^+\bar{K}^0/K^-K^0$ channels can be a good choice.  The typical decay, $\tau^- \to K^-K_S\nu_\tau$, has been measured
by CLEO~\cite{prd53-6037}, Belle~\cite{prd89-072009}, and {\it BABAR}~\cite{prd98-032010} Collaborations, and theoretical analyses
have been performed by including the contributions of the $\rho$ family~\cite{epjc39-41,mpla31-1650138,epjc79-436,epjconf212-03006}.
The pole mass of $m_{\rho(1700)}$ is close to the mass of the $\tau$ lepton; moreover, the full width for $\rho(1700)$
is broad~\cite{pdg2020}, so it is clearly unrealistic to find a complete curve of the intermediate resonance $\rho(1700)$ in the
$K^\pm K_S$ distribution in the $\tau$ decay. Within the QCD factorization approach,  the theoretical description for the decays
$B_c^+ \to J/\psi K^+K_S$ and $B_c^+ \to \psi(2S) K^+K_S$ including the resonances $\rho(1450)^+$ and $\rho(1700)^+$ has been
presented in Ref.~\cite{prd99-036019}. In this work, we shall study the charged $\rho(1450)$ and $\rho(1700)$ in the three-body
decays $B^+ \to \bar{D}^0 K^+ \bar{K}^0$, $B^0  \to D^- K^+ \bar{K}^0$, and $B_s^0 \to  D_s^-K^+ \bar{K}^0$ with the
corresponding quasi-two-body processes.

\begin{figure}[tbp]
\centerline{\epsfxsize=18 cm \epsffile{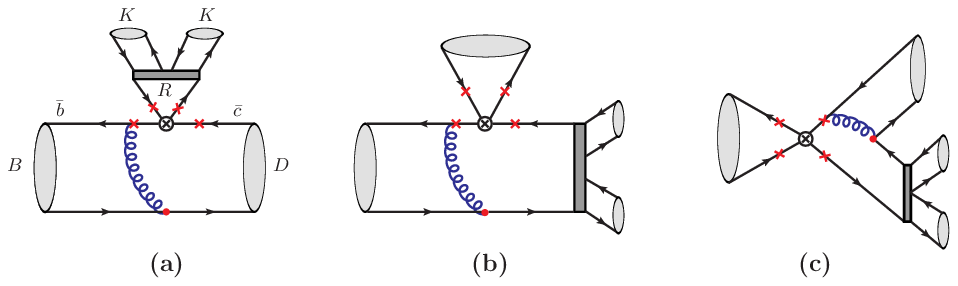} }
\vspace{-19cm}
\caption{Typical Feynman diagrams for the quasi-two-body decays $B\to DR \to D KK$, in which $D$ and $R$ represent
         the $D$ meson with $\bar{c}q$ and the intermediate resonances, respectively. The symbol $``\times"$ denotes the possible
         attachments for hard gluons,
         and $``\otimes"$ represents the insertion of the four-fermion vertices in the effective theory.
          }
\label{fig1}
\end{figure}

The decay $B_s^0\to  D_s^- R \to  D_s^-K^+ \bar{K}^0$
is contributed solely by the color-allowed emission diagrams as shown in Fig.~\ref{fig1}(a).
The decay amplitude is given by $\mathcal{A}_{B_s} \propto [(C_2+C_1/3)F_{eD}^{LL}+C_1M_{eD}^{LL}]$,
where $F_{eD}^{LL}$ and $M_{eD}^{LL}$ represent the amplitudes for factorizable and nonfactorizable parts respectively.
Evidently, the relevant factorizable amplitude is dominant and can be expressed as
$<K^+ \bar{K}^0|(\bar{u}d)_{V-A}|0><D_s^-|(\bar{b}c)_{V-A}|B_s^0>$.
Considering the $K^+ \bar{K}^0$ can only be produced by the vector current, the $\rho$ family resonances with $J^P=1^-$
should be prominent in the $KK$ system of $B_s^0 \to  D_s^-K^+ \bar{K}^0$.
For the other two decays, in addition to the color-allowed emission cases, the diagrams with the emission of a $D$ meson
and the annihilation diagrams as presented in Fig.~\ref{fig1} (b), (c) will also contribute to the decays $B^+ \to\bar{D}^0 R\to
\bar{D}^0 K^+ \bar{K}^0$, and $B^0  \to D^- R\to D^- K^+ \bar{K}^0$ respectively, but they are both color-suppressed.
It means that the decay amplitudes for these two decays should also be contributed mainly by the Feynman diagrams
with the resonances emitted, and the $\rho$ mesons can also be crucial in the $KK$ distribution of the corresponding
three-body $B$ decays. Moreover, the concerned quasi-two-body decays are CKM-favored processes, which should have relatively large
branching ratios and the potential to be measured by experiments.

In the past, the resonance contributions from various intermediate states for the three-body
decays $B \to D h_1h_2$ ($h_{1,2}$ represents pion or kaon) have been studied within the PQCD approach
~\cite{ijmpa35-2050164,prd102-056017,npb923-54,prd96-093011,epjc79-539,plb788-468,prd100-014017,
cpc43-073103,plb791-342,prd103-096016,2109.00664}.
In the recent work, we studied the virtual contributions for the kaon pair from the tail of $\rho(770)$
in the three-body decays $B \to D K\bar{K}$~\cite{prd103-016002}. Faced with the huge disparities between the coefficients for
the excited states of $\rho(770)$, $\omega(782)$, and $\phi(1020)$ in the kaon vector timelike form factor fitted in different 
studies~\cite{epjc39-41,prd81-094014,jetp129-386}, a detailed discussion was provided in Ref.~\cite{prd103-056021},
and the contributions for the kaon pair from $\rho(770,1450,1700)$ and $\omega(782,1420,1650)$  in the $B\to K\bar K h$ decays
have also been studied systematically.

The rest of this article is structured as follows.
In Sec.~\ref{sec-fra}, we provide a brief review of the PQCD framework for the concerned quasi-two-body decays.
In Sec.~\ref{sec-res}, we present the numerical results and the phenomenological analyses.
Finally, we provide a short summary in section~\ref{sec-con}.

\section{FRAMEWORK} \label{sec-fra}
Under the quasi-two-body approximation, the PQCD factorization formula for the decay amplitude ${\mathcal A}$ of the decay
$B\to D R \to D KK$ can be expressed as the convolution of a calculable hard kernel $H$ with the universal light-cone distribution
amplitudes~\cite{plb561-258,prd89-074031}
\begin{eqnarray} \label{def-DA}
{\cal A} \sim \Phi_B\otimes H\otimes \Phi_{D}\otimes\Phi_{KK},
\end{eqnarray}
in which $\Phi_B$ and $\Phi_{D}$ are the distribution amplitudes for the $B$ and $D$ mesons, respectively. The  distribution amplitude
$\Phi_{KK}$ is introduced to describe the interactions between the kaon pair originating from the resonance $R$.

In the PQCD approach, the distribution amplitudes for the initial and final states are
crucial inputs. In this work, we employ the same forms of $\Phi_B$ and $\Phi_{D}$ as those widely
adopted in PQCD; one can find their explicit expressions and parameters in Ref.~\cite{npb923-54} and the references therein.
For the $P$-wave $KK$ system along with the subprocesses $\rho^+ \to KK$, the distribution amplitudes are organized into
\cite{prd101-111901,epjc80-815}
\begin{eqnarray}\label{def-DA-KK}
     \phi^{P\text{-wave}}_{KK}(z,s)
             =\frac{-1}{\sqrt{2N_c}} \left[\sqrt{s}\,{\epsilon\hspace{-1.5truemm}/}\!_L\phi^0(z,s)
                      +  {\epsilon\hspace{-1.5truemm}/}\!_L {p\hspace{-1.7truemm}/} \phi^t(z,s)+\sqrt s \phi^s(z,s)  \right],
\end{eqnarray}
where $z$ is the momentum fraction for the spectator quark, $s$ is the squared invariant mass of the
kaon pair, and $\epsilon_L$ and $p$ are the longitudinal polarization vector and momentum for the resonances, respectively.
The twist-2 and twist-3 distribution amplitudes $\phi^{0}$, $\phi^{s}$, and $\phi^{t}$ are chosen as
\begin{eqnarray}
   \phi^{0}(z,s)&=&\frac{3 F_K(s)}{\sqrt{2N_c}} z(1-z)\left[1+a_2^{0} C^{3/2}_2(1-2z) \right]\!,\label{def-DA-0}\\
  \phi^{s}(z,s)&=&\frac{3 F^s_K(s)}{2\sqrt{2N_c}}(1-2z)\left[1+a_2^s\left(1-10z+10z^2\right) \right]\!,\label{def-DA-s}\\
   \phi^{t}(z,s)&=&\frac{3 F^t_K(s)}{2\sqrt{2N_c}}(1-2z)^2\left[1+a_2^t  C^{3/2}_2(1-2z)\right]\!.\label{def-DA-t}\quad
\end{eqnarray}
with the Gegenbauer polynomial $C^{3/2}_2(t)=3\left(5t^2-1\right)/2$ and
the  Gegenbauer moments $a_2^{0}=0.25\pm0.10$, $a_2^s=0.75\pm0.25$, and $a_2^t=-0.60\pm0.20$~\cite{plb763-29,prd101-111901}.
The vector time-like form factors $F^s_K(s)$ and $F^t_K(s)$ for the twist-$3$ distribution amplitudes are deduced by
the assumption $F^{s,t}_K(s)\approx (f^T_{\rho}/f_{\rho})F_K(s)$~\cite{plb763-29,prd103-056021} with
$f_\rho=0.216$ GeV and $f^T_\rho=0.165$ GeV~\cite{prd75-054004}.

For the $P$-wave $K^+K^-$ and $K^0\bar{K}^0$ systems, the kaon form factors are generally
saturated by $\rho(770)$, $\omega(782)$, $\phi(1020)$, and their radial excitations.
When it comes to the $K^+ \bar{K}^0$ case, only the $\rho$ family states contribute, and
the vector timelike form factors can be simply expressed as~\cite{epjc39-41,prd103-016002,prd103-056021}
\begin{eqnarray}
  F_K(s)&=&\sum_{\rho_i} c^K_{\rho_i} {\rm BW}_{\rho_i}(s),
  \label{def-F-uds}
\end{eqnarray}
where the symbol $\rho_i$ indicates the resonance $\rho(770)$ and its excited states.
For the normalization factors for $\rho(1450)$ and $\rho(1700)$, we take $c^K_{\rho(1450)}=-0.156\pm0.015$
and $c^K_{\rho(1700)}=-0.083\pm0.019$ as discussed in detail in Ref.~\cite{prd103-056021}.
The resonance shape for $\rho(1450,1700)$ is described by the KS version of the BW formula~\cite{zpc48-445,epjc39-41}:
\begin{eqnarray}\label{def-BW}
   {\rm BW}_R(s) = \frac{m_R^2}{m_R^2-s-i\sqrt{s}\Gamma_{tot}(s)}\,.    \label{BW}
\end{eqnarray}
where the mass-dependent width $\Gamma_{tot}(s)$ can be found in Refs.~\cite{jhep01-112,2111.12307}.

By combining various of contributions from the relevant Feynman diagrams as shown in Fig.~\ref{fig1},
the total decay amplitudes for the concerned quasi-two-body decays are written as
\begin{eqnarray}\label{DA}
   \mathcal{A}({B^+ \to \bar{D}^0} [\rho^+\to] K^+ \bar{K}^0)&=&\frac{G_F}{\sqrt2}V^*_{cb}V_{ud}
   [a_2F_{eD}^{LL}+C_1M_{eD}^{LL}+a_1F_{e\rho}^{LL}+C_2
   M_{e\rho}^{LL} ], \\
   \mathcal{A}({B^0 \to {D^-} [\rho^+  \to]  K^+ \bar{K}^0})&=&\frac{G_F}{\sqrt2}V^*_{cb}V_{ud}
   [a_2F_{eD}^{LL}+C_1M_{eD}^{LL} +a_1F_{a\rho}^{LL}+C_2M_{a\rho}^{LL}], \\
   \mathcal{A}({B_s^0 \to {D_s^-} [\rho^+ \to]  K^+ \bar{K}^0})&=&\frac{G_F}{\sqrt2}V^*_{cb}V_{ud}
   [a_2F_{eD}^{LL}+C_1M_{eD}^{LL} ],
\end{eqnarray}
where $G_F = 1.16638 \times 10^{-5}~{\rm GeV}^{-2}$ is the Fermi coupling constant, $V_{cb,ud}$ are the CKM matrix elements, and
$a_1=C_1+ C_2/3$ and $a_2=C_2+ C_1/3$ are the Wilson coefficients. The PQCD calculations of individual amplitudes
$F^{LL}_{eD,e\rho,a\rho}$ and $M^{LL}_{eD,e\rho,a\rho}$ for the factorizable and nonfactorizable Feynman diagrams
are carried out in the frame with the $B$ meson at rest. Because of the same quark composition, the decay amplitudes
for the considered decays with the intermediate resonances $\rho(1450,1700)$ are the same as those with $\rho(770)$, and
the explicit expressions, together with the definitions of related momenta in the light-cone coordinates,
have been presented in Ref.~\cite{prd103-016002}. For the sake of simplicity, we shall not present them in this
work.

\section{RESULTS} \label{sec-res}
The masses for the initial and final particles, the full widths for the resonance states, and
the decay constants for $B$ and $D$ mesons are summarized as follows (in units of GeV)~\cite{pdg2020}:
\begin{eqnarray}
 m_{B^\pm/B^0/B_s^0}&=&5.279/5.280/5.367,\quad m_{D^\pm/D^0/D_s^\pm}=1.870/1.865/1.968, \nonumber\\
 \quad m_{K^\pm/K^0}&=&0.494/0.498, \qquad m_{\rho(1450)}=1.465, \qquad m_{\rho(1700)}=1.720,\nonumber\\
 \Gamma_{\rho(1450)}&=&0.400\pm0.060, \qquad \Gamma_{\rho(1700)}=0.250\pm0.100,\nonumber\\
\quad f_{B^{\pm,0}/B_s^0}&=&0.190/0.230,\qquad f_{D^{\pm,0}/D_s^\pm}=0.213/0.250. \nonumber \label{eq:inputs}
\end{eqnarray}
In addition, the mean lifetimes $\tau_{B^\pm/B^0/B^0_s}=1.638/1.519/1.515$~ps, and the Wolfenstein parameters for the CKM matrix
$A=0.790^{+0.017}_{-0.012}$, $\lambda=0.22650\pm0.00048$, $\bar{\rho} = 0.141^{+0.016}_{-0.017}$, and $\bar{\eta}= 0.357\pm0.011$
are also adopted as in Ref.~\cite{pdg2020}.

For the considered $B \to D \rho(1450)^+ \to D K^+ \bar{K}^0$ decays,
the PQCD predictions for the branching ratios of each case are the following:
\begin{eqnarray}\label{BR1450}
   \mathcal{B}(B^+ \to {\bar{D}^0} [\rho(1450)^+\to] K^+ \bar{K}^0)&=&8.99^{+5.17}_{-3.07}(\omega_B)
    ^{+0.57}_{-0.52}(C_D)^{+0.57}_{-0.56}(a^{0,s,t}_2)^{+1.81}_{-1.65}(c^K_{\rho(1450)})\times 10^{-5},\nonumber \\
   \mathcal{B}(B^0 \to {D^-} [\rho(1450)^+  \to] K^+ \bar{K}^0)&=&5.02^{+3.34}_{-1.89}(\omega_B)
    ^{+0.46}_{-0.45}(C_D)^{+0.39}_{-0.34}(a^{0,s,t}_2)^{+1.01}_{-0.92}(c^K_{\rho(1450)})\times 10^{-5},\nonumber \\
   \mathcal{B}(B_s^0 \to {D_s^-} [\rho(1450)^+ \to] K^+ \bar{K}^0)&=&4.37^{+2.75}_{-1.56}(\omega_{B_s})
    ^{+0.34}_{-0.33}(C_{D_s})^{+0.03}_{-0.03}(a^{0,s,t}_2)^{+0.88}_{-0.80}(c^K_{\rho(1450)})\times 10^{-5},  \nonumber\\
\end{eqnarray}
while the predicted branching ratios of same decays with the intermediate resonance $\rho(1700)$  are
\begin{eqnarray}\label{BR1700}
   \mathcal{B}(B^+ \to {\bar{D}^0} [\rho(1700)^+\to] K^+ \bar{K}^0)&=&8.28^{+4.60}_{-2.81}(\omega_B)
    ^{+0.59}_{-0.46}(C_D)^{+0.61}_{-0.57}(a^{0,s,t}_2)^{+4.42}_{-3.35}(c^K_{\rho(1700)}) \times 10^{-5}, \nonumber \\
   \mathcal{B}(B^0 \to {D^-} [\rho(1700)^+  \to] K^+ \bar{K}^0)&=&4.00^{+2.60}_{-1.50}(\omega_B)
    ^{+0.38}_{-0.31}(C_D)^{+0.46}_{-0.34}(a^{0,s,t}_2)^{+2.04}_{-1.62}(c^K_{\rho(1700)})\times 10^{-5},\nonumber \\
   \mathcal{B}(B_s^0 \to {D_s^-} [\rho(1700)^+ \to] K^+ \bar{K}^0)&=&3.88^{+2.41}_{-1.38}(\omega_{B_s})
    ^{+0.29}_{-0.28}(C_{D_s})^{+0.03}_{-0.03}(a^{0,s,t}_2)^{+1.98}_{-1.57}(c^K_{\rho(1700)})\times 10^{-5}.\nonumber \\
\end{eqnarray}
All the results above are given in the energy range with the invariant mass of the $K^+ \bar{K}^0$ pair
varying between $m_{K^+}+m_{\bar{K}^0}$ and $m_B-m_D$. The first two theoretical errors come from the uncertainties of
the parameters $\omega_{B_{(s)}} = 0.40 \pm 0.04~(0.50 \pm 0.05)$ and $C_{D_{(s)}}=0.5\pm 0.1~(0.4\pm0.1)$ in the
distribution amplitudes for the $B$ and $D$ mesons, respectively; the errors from the uncertainties of Gegenbauer moments
$a^0_2=0.25\pm0.10$, $a^s_2=0.75\pm0.25$, and $a^t_2= -0.60\pm0.20$ in the two-kaon system are added together as the third error;
and the last one is due to the coefficient $c^K_{\rho(1450)}=-0.156\pm0.015$
or $c^K_{\rho(1700)}=-0.083\pm0.019$ in the kaon timelike form factor. The errors from the uncertainties of
other parameters are relatively small and have been neglected.

From the calculations and the numerical results, one can conclude the following:
\begin{itemize}
\item[(1)]
The branching ratios of the considered quasi-two-body decays, which are expected to be large
and measurable, are predicted to be in the order of $10^{-5}$ numerically in the PQCD approach.
We also test the contributions from the color-suppressed emission diagrams
 for $B^+ \to {\bar{D}^0} [\rho(1450)^+\to] K^+ \bar{K}^0$ and the annihilation diagrams
 for $B^0 \to {D^-} [\rho(1450)^+  \to] K^+ \bar{K}^0$, and the corresponding branching ratios
 are obtained to be $2.18 \times 10^{-6}$
 and $1.06 \times 10^{-6}$, while the results become $6.92\times 10^{-5}$ and $6.45\times 10^{-5}$, respectively,
when considering only the color-allowed emission diagrams for those two decays. Although
the interference between the different types of Feynman diagrams for each decay
 are non-negligible, the contributions from the diagrams with the emission of a kaon pair are still dominant
 as expected.

\item[(2)]
The PQCD predicted branching ratios of the decay modes with the subprocess  $\rho(1700)^+\to K^+ \bar{K}^0$  are not much smaller than
the corresponding ones with the intermediate state $\rho(1450)^+$. This is because the latter are suppressed in the phase space
when considering the mass of the kaon pair $m_{K^+}+m_{\bar{K}^0}$ is very close to $m_{\rho(1450)}-\Gamma_{\rho(1450)}$ but considerably
less than $m_{\rho(1700)}-\Gamma_{\rho(1700)}$ in comparison. In contrast, the situation is quite different in the same $B$ meson
decays with the subprocesses $\rho(1450) \to \pi\pi$ and $\rho(1700) \to \pi\pi$
~\cite{prd96-036014,prd96-093011,prd98-113003,prd90-012003,plb742-38,prd92-032002} on account of
 the impact on the phase space arising from the difference between the mass of the pion and kaon.

\item[(3)] {
In the measurement by the Belle Collaboration~\cite{plb542-171}, the branching fraction of
the $B^+ \to \bar{D}^0 K^+ \bar{K}^0$ decay was determined to be
$(5.5\pm1.4\pm 0.8)\times 10^{-4}$. Meanwhile, they also gave a branching ratio for $B^0\to D^-K^+ \bar{K}^0$
of $(1.6\pm0.8\pm 0.3)\times 10^{-4}$ and set an upper limit as $3.1\times 10^{-4}$ at $90\%$
confidence level.
With the center values of our predictions in Eqs.~(\ref{BR1450})-(\ref{BR1700}),
 we obtain the ratios
\begin{eqnarray}\label{R1}
\frac{\mathcal{B}(B^+ \to \bar{D}^0 \rho(1450)^+ [\to K^+ \bar{K}^0])}
{\mathcal{B}(B^+ \to \bar{D}^0 K^+ \bar{K}^0)}=16.35\%
\end{eqnarray}
 and
 \begin{eqnarray}\label{R2}
 \frac{\mathcal{B}(B^+ \to \bar{D}^0 \rho(1700)^+ [\to K^+ \bar{K}^0])}
{\mathcal{B}(B^+ \to \bar{D}^0 K^+ \bar{K}^0)}=15.05\%.
\end{eqnarray}

Similarly, the ratios become $31.38\%$ and $25.00\%$ for the $B^0  \to D^- K^+ \bar{K}^0$ decay mode.
 Since there are still large uncertainties in the branching fraction of
 $B^0  \to D^- K^+ \bar{K}^0$ presented by Belle, more precise measurements for this three-body channel are needed.
Nevertheless, the above results indicate that the contributions from charged $\rho(1450)$ and $\rho(1700)$ can
be important in the relevant three-body $B$ meson decays.}

\item[(4)]
In our previous work~\cite{prd103-016002}, the virtual contribution for $\rho(770)$ in $B\to DKK$ were studied, and
the center values of the branching ratios for the decays $B^+ \to {\bar{D^0}} \rho(770)^+$,
 $B^0 \to {D^-} \rho(770)^+$, and $B_s^0 \to {D_s^-} \rho(770)^+$ with the subprocess $\rho(770)^+\to K^+ \bar{K}^0$
 are predicted to be $1.18 \times 10^{-4}$, $7.93 \times 10^{-5}$, and $6.06 \times 10^{-5}$, respectively.
 Comparing those results with the numerical results in this study, two conclusions can be obtained. On the one hand, the contribution
 from the tail of $\rho(770)$ is still large and cannot be ignored in the $P$-wave $K^+ \bar{K}^0$ system.
 On the other hand,
 limited by the phase space for $KK$ production in the decay of the $\rho(770)$ meson,
the contributions for $KK$ pair from the excited states $\rho(1450)$ and $\rho(1700)$
 are more close to that of $\rho(770)$ compared with the corresponding ones in the $\pi\pi$ system.
 Therefore, it is possible to find the resonances $\rho^+(1450)$ and $\rho^+(1700)$
 in the $KK$ channel, and the concerned $B\to DKK$ decays can be employed to study the properties of those excited $\rho$ mesons.
 Interestingly, the $KK_S$ invariant mass spectrum for the $B^+ \to \bar{D}^0 K^+ K^0_S$ decay presented
  by Belle showed a peak around $1.2$ GeV~\cite{plb542-171}. It indicates that other possible resonances,
  such as $a_0(980)$ and $a_2(1320)$,
  will contribute to this region in addition to $\rho(770)$ and $\rho(1450)$. Moreover, the interference between $\rho(770)$ and
  $\rho(1450)$ can increase or decrease the total contributions from these two resonances. We leave these possible contributions
  for the kaon pair to future studies due to the lack of precise parameters of the kaon time-like form factor,
  and more accurate measurements for this decay will help to test the $KK$ distribution close to the threshold determined by Belle.
\end{itemize}

\section{Summary} \label{sec-con}
In this work, we studied the resonance contributions for $\rho(1450)^+$ and $\rho(1700)^+$
in the $B^+ \to \bar{D}^0 K^+ \bar{K}^0$, $B^0  \to D^- K^+ \bar{K}^0$, and $B_s^0 \to  D_s^-K^+ \bar{K}^0$ decays.
The branching ratios of the concerned quasi-two-body $B$ meson decays are predicted to be in the order of $10^{-5}$
numerically within the PQCD approach. The contributions from the intermediate states
$\rho(1450)$ and $\rho(1700)$ are $16.35\%$ and $15.05\%$ of the data for the $B^+ \to \bar{D}^0 K^+ \bar{K}^0$ decay presented by Belle.
The similar ratios become $31.38\%$ and $25.00\%$ for $B^0  \to D^- K^+ \bar{K}^0$, where the total three-body
branching fraction
still has large errors and needs further measurements.
It shows that the contributions from charged $\rho(1450)$ and $\rho(1700)$ can be important
in the relevant three-body $B$ meson decays, and
all these PQCD predictions will be tested in the LHCb and Belle-II experiments in the future.

\begin{acknowledgments}
This work was supported by the National Natural Science Foundation of China under Grant No.~11947011,
 the Natural Science Foundation of Jiangsu Province under Grant
No.~BK20191010 and the Scientific Research Foundation of Nanjing Institute of Technology under Grant No.~YKJ201854.
\end{acknowledgments}


\end{document}